\documentclass[aps,pra,twocolumn,amsfonts,amssymb,amsmath,showpacs,
floatfix,nofootinbib,groupedaddress,superscriptaddress,citesort]{revtex4}
\usepackage{mathrsfs}
\usepackage{amsfonts}
\usepackage{amstext}
\usepackage{amsmath}
\usepackage{amssymb}
\usepackage{bm}
\usepackage{bbm}
\usepackage[dvips]{graphicx}
\def\qed{\leavevmode\unskip\penalty9999 \hbox{}\nobreak\hfill
     \quad\hbox{\leavevmode  \hbox to.77778em{%
              \hfil\vrule   \vbox to.675em%
               {\hrule width.6em\vfil\hrule}\vrule\hfil}}
     \par\vskip3pt}

\begin{document}

\title{Conditions for coherence transformations under incoherent operations}

\author{Shuanping Du}
\email{shuanpingdu@yahoo.com} \affiliation{School of Mathematical
Sciences, Xiamen University, Xiamen, Fujian, 361000, China}

\author{Zhaofang Bai}\thanks{Corresponding author}
\email{baizhaofang@xmu.edu.cn} \affiliation{School of Mathematical
Sciences, Xiamen University, Xiamen, Fujian, 361000, China}

\author{Yu Guo}
\email{guoyu3@aliyun.com} \affiliation{School of Mathematics and
Computer Science, Shanxi Datong University, Datong 037009, China}

\begin{abstract}

We build the counterpart of the celebrated Nielsen's theorem for
coherence manipulation in this paper. This offers an affirmative
answer to the open question: whether, given two states $\rho$ and
$\sigma$, either $\rho$ can be transformed into $\sigma$ or vice
versa under incoherent operations [Phys. Rev. Lett. \textbf{113},
140401(2014)]. As a consequence, we find that there exist
essentially different types of coherence. Moreover, incoherent
operations can be enhanced in the presence of certain coherent
states. These extra states are coherent catalysts: they allow
uncertain incoherent operations to be realized, without being
consumed in any way. Our main result also sheds a new light on the
construction of coherence measures.

\end{abstract}

\pacs{03.65.Ud, 03.67.-a, 03.65.Ta.}
\maketitle

{\it Introduction.---} Superposition is a critical property of
quantum system resulting in quantum coherence and quantum
entanglement. Quantum coherence and also entanglement provide the
important resource for quantum information processing, for
example, Deutsch¡¯s algorithm, Shor¡¯s algorithm, teleportation,
superdense coding and quantum cryptography \cite{Nielsen}. As with
any such resource, there arises naturally the question of how it
can be quantified and manipulated. Attempts have been made to find
meaningful measures of entanglement
\cite{Ben1,Ben2,Ben3,Ved1,Ved2}, and also to uncover the
fundamental laws of its behavior under local quantum operations
and classical communication (LOCC)
\cite{Ben1,Ben2,Ben3,Ved1,Ved2,Nie2,Vid, Jon1,Jon2,Har,Nie3}. The
celebrated Nielsen's theorem finds possible entanglement
manipulation between bipartite entanglement states by LOCC
\cite{Nie2}. Let $|\psi\rangle=\sum_{i=1}^d \sqrt
{\psi_j}|jj\rangle$ and $|\phi\rangle=\sum_{i=1}^d \sqrt
{\phi_j}|jj\rangle$ be two bipartite states whose Schmidt
coefficients are ordered in decreasing order,
$\psi_1\geq\psi_2\geq\cdots\geq\psi_d$,
$\phi_1\geq\phi_2\geq\cdots\geq\phi_d$. Then
$|\psi\rangle\rightarrow|\phi\rangle$ by LOCC if and only if
$(\psi_1,\psi_2,\cdots,\psi_d)\prec
(\phi_1,\phi_2,\cdots,\phi_d)$. This reveals a partial ordering on
the entangled states and connects quantum entanglement to the
algebraic theory of majorization.

In \cite{Bau}, the researchers establish a rigorous framework for
the quantification of coherence as a resource following the
viewpoints that have been established for entanglement in
\cite{Ved2}. And the setting of single copies of coherent states
is of considerable interest from the practical point of view as
this is most readily accessible in the laboratory.  It is expected
that theory of coherence manipulation can be established that
proceeds along analogous developments in entanglement theory
\cite{Bau}. The aim of this paper is to build the counterpart of
the Nielsen's theorem for coherence manipulation. What is amazing
is that majorization is also the key ingredient. It provides the
relevant structure that determines the interconvertibility of
coherent states.

 Majorization is an active research area in linear algebra.
We use Chap. 2 of \cite{Bha} as our principal reference on
majorization. Suppose $x=(x_1,x_2,\cdots, x_d)^t$ and
$y=(y_1,y_2,\cdots ,y_d)^t$ are real d-dimensional vectors, here
$x=(x_1,x_2,\cdots, x_d)^t$ denotes the transpose of row vector
$(x_1,x_2,\cdots, x_d)$. Then $x$ is majorized by $y$
(equivalently $y$ majorizes $x$), written $x\prec y$, if for each
$k$ in the range $1, \cdots, d$, $\sum_{i=1}^k x_i^\downarrow\leq
\sum_{i=1}^k y_i^\downarrow$ with equality holding when $k= d$,
and where the $x_i^\downarrow$ indicates that elements are to be
taken in descending order, so, for example, $x_1^\downarrow$ is
the largest element in $(x_1,\cdots , x_d)$. The majorization
relation is a partial order on real vectors, with $x \prec y$ and
$y \prec x$ if and only if $x^\downarrow=y^\downarrow$.

In the following, we introduce the concepts of incoherent states
and incoherent operations which are from \cite{Bau}. Let
${\mathcal H}$ be a finite dimensional Hilbert space with
$\dim({\mathcal H})=d$. Fixing a particular basis
$\{|i\rangle\}_{i=1}^d$, we call all density operators (quantum
states) that are diagonal in this basis incoherent, and this set
of quantum states will be labelled by ${\mathcal I}$, all density
operators $\rho\in {\mathcal I}$ are of the form
\begin{equation}\rho=\sum_{i=1}^d\lambda_i|i\rangle\langle i|. \end{equation} Quantum
operations are specified by a finite set of Kraus operators
$\{K_n\}$ satisfying $\sum_n K_n^\dag K_n=I$, $I$ is the identity
operator on ${\mathcal H}$. Quantum operations are incoherent if
they fulfil $K_n\rho K_n^\dag/Tr(K_n\rho K_n^\dag)\in {\mathcal
I}$ for all $\rho\in {\mathcal I}$ and for all $n$.

{\it Results.--- }To state our central result linking coherence
manipulation with majorization, we need some notation. Suppose
$|\psi\rangle=\sum_{i=1}^d\psi_i|i\rangle$ and
$|\phi\rangle=\sum_{i=1}^d\phi_i|i\rangle$ are any pure states.
$|\psi\rangle \xrightarrow{ICO}|\phi\rangle$, read
``$|\psi\rangle$ transforms incoherently to$|\phi\rangle$"
indicates that $|\psi\rangle\langle\psi|$ transforms to
$|\phi\rangle\langle\phi|$ by incoherent operations. Then we have
the following:

\textbf {Theorem 1.} $|\psi\rangle$  transforms  to $|\phi\rangle$
using incoherent operations if and only if $(|\psi_1|^2,\cdots,
|\psi_d|^2)^t$ is majorized by $(|\phi_1|^2,\cdots,
|\phi_d|^2)^t$. More succinctly,
\begin{equation}\begin{array}{l}|\psi\rangle \xrightarrow{ICO}|\phi\rangle
\quad \text{iff} \\
 (|\psi_1|^2,\cdots, |\psi_d|^2)^t\prec(|\phi_1|^2,\cdots,
|\phi_d|^2)^t.\end{array}\end{equation}


One direct consequence of Theorem 1 is that there exist pairs
$|\psi\rangle$ and $|\phi\rangle$  with neither $|\psi\rangle
\xrightarrow{ICO}|\phi\rangle$ nor $|\phi\rangle
\xrightarrow{ICO}|\psi\rangle$. For example, $d=3$,
\begin{equation}|\psi\rangle=\sqrt{0.4}|1\rangle+\sqrt{0.3}|2\rangle+\sqrt{0.3}|3\rangle,\end{equation}
\begin{equation}|\phi\rangle=\sqrt{0.5}|1\rangle+\sqrt{0.1}|2\rangle+\sqrt{0.4}|3\rangle.\end{equation}
These provide  an example of essentially different types of
coherence, from the point of view of incoherent operations. We
will say that $|\psi\rangle$ and $|\phi\rangle$ are incomparable
in coherence. In addition, for any two pure states $|\psi\rangle$,
$|\phi\rangle$, $|\psi\rangle$ and $|\phi\rangle$ can be
incomparable with respect to incoherence under a change of basis.
This may seem odd at first, but it turns out that coherence is a
basis dependent phenomenon.

For entanglement transformations, a major interest  has been the
catalysis. This enables the conversion between two initially
inconvertible entangled states assisted by a lent entangled state,
which is recovered at the end of the process
\cite{Jon2,Ghi,Son,Cha,Bow,Ber}. For two states $|\psi\rangle$ and
$|\phi\rangle$ which are incomparable in coherence, if
$|\psi\rangle|\delta\rangle
\xrightarrow{ICO}|\phi\rangle|\delta\rangle$, we say $|\psi\rangle
$ is transformed into $|\phi\rangle$ under coherence-assisted
incoherent operation, and $|\delta\rangle$ is called a coherent
catalyst. This state acts  much like a catalyst in a chemical
reaction: its presence allows a previously forbidden
transformation to be realized, and since it is not consumed it can
be reused. Here we use the phrase ``coherence-assisted"  because
$|\delta\rangle$ must be coherent. Combining Theorem 1 and proofs
of Lemma 1, Lemma 2 and Lemma 3 in \cite{Jon2}, we immediately
have the following interesting results:

(i) No incoherent transformation can be catalyzed by a maximally
coherent state $|\psi_d\rangle=\sum_{k=1}^d \frac 1 {\sqrt{d}}
|k\rangle $. This shows a surprising property of coherent
catalysts: they must be partially coherent. If the catalyst has
not enough coherence, then $|\psi\rangle$ can not be transformed
into $|\phi\rangle$ with certainty, but if it has too much then
the result is same.

(ii) Two states are interconvertible (i.e., both
$|\psi\rangle\rightarrow |\phi\rangle$ and
$|\phi\rangle\rightarrow |\psi\rangle$) under coherence-assisted
incoherent operation if and only if they are equivalent up to a
permutation of diagonal  unitary transformations. One consequence
of this result is that if a transition that is forbidden under
incoherent operation can be catalyzed (i.e.
$|\psi\rangle\nrightarrow |\phi\rangle$ under incoherent operation
but $|\psi\rangle|\delta\rangle\rightarrow
|\phi\rangle|\delta\rangle$ for some $|\delta\rangle$), then the
reverse transition (from $|\phi\rangle\rightarrow |\psi\rangle$
can not be catalyzed. In particular, only transitions between
incomparable states may be catalyzed.

(iii) $|\psi\rangle\rightarrow |\phi\rangle$ under
coherence-assisted incoherent operation only if both
$|\psi_1|\leq|\phi_1|$ and $|\psi_d|\geq|\phi_d|$.

Theorem 1 provides a necessary condition for coherence measures.
By \cite{Bau}, coherence measures should satisfy the monotonicity
under incoherent operations, i.e., ${\mathcal
C}(\Phi(\rho))\leq{\mathcal C}(\rho)$ for any incoherent operation
$\Phi$ and state $\rho$. Let
$|\psi\rangle=\psi_1|1\rangle+\cdots+\psi_d|d\rangle$,
$|\phi\rangle=\phi_1|1\rangle+\cdots+\phi_d|d\rangle$, with
$(|\psi_1|^2,\cdots, |\psi_d|^2)^t\prec (|\phi_1|^2,\cdots,
|\phi_d|^2)^t$. By Theorem 1, we have ${\mathcal
C}(|\phi\rangle\langle\phi|)\leq{\mathcal
C}(|\psi\rangle\langle\psi|)$. This necessary condition of
coherence measure implies the Result 1 in \cite{Gir} is not true.
That is, Wigner-Yanase-Dyson skew information
\begin{equation}{\mathcal C}(\rho, K)=-\frac{1}{2}Tr([\sqrt{\rho},K]^2)\end{equation} is not
a good coherence measure since it violates this necessary
condition. Assume $d=3$, let\begin{equation}\begin{array} {ll}
K&=|1\rangle\langle 1|+10|2\rangle\langle
2|+5|3\rangle\langle 3|,\\
|\psi\rangle&=\frac{1}{\sqrt 3}|1\rangle+\frac{1}{\sqrt
3}|2\rangle+\frac{1}{\sqrt 3}|3\rangle,\\
|\phi\rangle&=\frac{1}{\sqrt 2}|1\rangle+\frac{1}{\sqrt
2}|2\rangle.\end{array}\end{equation} It is easy to check that
$(\frac{1}{3}, \frac{1}{3},\frac{1}{3})^t\prec (\frac{1}{2},
\frac{1}{2},0)^t$ and \begin{equation}{\mathcal
C}(|\phi\rangle\langle \phi|,K)=\frac{81}{4}> {\mathcal
C}(|\psi\rangle\langle \psi|,K)=\frac{122}{9}.\end{equation}

The following construction of coherent measures is originated from
Theorem 1. For arbitrary pure state $|\psi\rangle=\sum_{i=1}^d
\psi_i |i\rangle$, we define $C_l(|\psi\rangle\langle
\psi|)=\sum_{i=l}^d {|\psi_i|^2}^{\downarrow}$ ($l=2,3,\cdots,
d$), here
$({|\psi_1|^2}^{\downarrow},{|\psi_2|^2}^{\downarrow},\cdots,{|\psi_d|^2}^{\downarrow})^t$
is the vector obtained by rearranging the coordinates of
$(|\psi_1|^2,|\psi_2|^2,\cdots,|\psi_d|^2)^t$ in the decreasing
order, and extending it over the whole set of density matrices as
$ C_l(\rho)=\min _{p_j,\rho_j}\sum_j p_jC_l(\rho_j),$ where the
minimization is to be performed over all the pure-state ensembles
of $\rho$, i.e., $\rho=\sum_j p_j\rho_j$. In \cite{BDQ}, we show
that $C_l$ are coherence measures.

Theorem 1 also pave the way for the following question: suppose
there is a pure coherent state
$|\psi\rangle=\sum_{i=1}^d\psi_i|i\rangle$ and we would like to
convert it into another pure coherent state
$|\phi\rangle=\sum_{i=1}^d\phi_i|i\rangle$ by incoherent
operations. Which is the greatest probability of success in such a
conversion? In \cite{BDQ}, we give the explicit formula of the
greatest probability
$P(|\psi\rangle\xrightarrow{ICO}|\phi\rangle)$. A parallel result
in entanglement theory is optimal local conversion strategy
between any two pure entangled states of a bipartite system
\cite{Vid}.

{\it Proofs.---} Now we do some preparatory work to prove Theorem
1 by collecting some useful facts:

(i) For real vectors $x,y$, $x\prec y$ if and only if $x=A y$ for
some doubly stochastic matrix. Recall that a $d\times d$ matrix $A
= ( a_{ij} )$ is called doubly stochastic if $a_{ij}\geq 0$ and
$\sum_{i=1}^d a_{ij}=\sum_{j=1}^d a_{ij}=1$.

(ii) For every  doubly stochastic matrix $A$, it  is a matrix that
may be written as a product of at most $d-1$ $T-$transforms. A
$T-$ transform, by definition, acts as the identity on all but two
matrix components. On those two components, it has the form
\begin{equation}T=\left(\begin{array}{cc} t& 1-t\\1-t & t\end{array}\right),\end{equation}
where $0\leq t\leq 1$. In terms of transformation,
$T(x_1,x_2,\cdots,
x_d)^t=(x_1,\cdots,x_{i-1},tx_i+(1-t)x_j,x_{i+1},\cdots
x_{j-1},(1-t)x_i+tx_j,x_{j+1},\cdots,x_d)^t$ for some indies $i,j$
and $0\leq t\leq 1$.\vspace{0.1in}

(iii) Let $\pi$ be a permutation of $\{1,2,\cdots,d\}$ and $P_\pi$
be the permutation matrix corresponding to $\pi$ which  is
obtained by permuting the rows  of a $d\times d$ identity matrix
according to $\pi$. A permutation matrix has exactly one entry 1
in each row and each column and 0 elsewhere.

(iv) For quantum operation $\Phi(\cdot)=\sum_n K_n\cdot K_n^\dag$,
it is easy to see that $\Phi$ is incoherent if and only if every
column of $K_n$ in the fixed basis $\{|i\rangle\}_{i=1}^d$ is with
at most 1 nonzero entry.

Now, we are in the position to give the proof of Theorem 1.

Proof: Firstly, we can suppose all  $\psi_k,\phi_k$
$(k=1,2,\cdots,d)$ are nonnegative and sorted in descending order.
Indeed, in general case, let $\psi_k=|\psi_k|e^{i\alpha_k}$,
$\phi_k=|\phi_k|e^{i\beta_k}$ and $|\psi_{\pi(1)}|\geq
|\psi_{\pi(2)}|\geq \cdots\geq |\psi_{\pi(d)}|$,
$|\phi_{\sigma(1)}|\geq |\phi_{\sigma(2)}|\geq \cdots\geq
|\phi_{\sigma(d)}|$, where $\pi,\sigma$ are two permutations of
$\{1,2,\cdots,d\}$. One can  define
$U=P_{\pi}\text{diag}(e^{-i\alpha_1},
e^{-i\alpha_2},\cdots,e^{-i\alpha_d})$ and
$V=P_{\sigma}\text{diag}(e^{-i\beta_1},
e^{-i\beta_2},\cdots,e^{-i\beta_d})$, here $P_{\pi}$ and
$P_{\sigma}$ are permutation matrices corresponding to $\pi$ and
$\sigma$, respectively. Note that $U|\psi\rangle
\xrightarrow{ICO}V|\phi\rangle\Leftrightarrow|\psi\rangle
\xrightarrow{ICO}|\phi\rangle$, we can replace $|\psi\rangle$ and
$|\phi\rangle$ by $U|\psi\rangle$ and $V|\phi\rangle$.

Now, we prove the ``if" part. Assume that $(|\psi_1|^2,\cdots,
|\psi_d|^2)^t\prec(|\phi_1|^2,\cdots, |\phi_d|^2)^t$. We will
apply the inductive method.

Assume  $\dim H=2$. If $\psi_2=0$, from the majorization, it
follows that $\phi_2=0$. That is $|\psi\rangle=|\phi\rangle
=|1\rangle $. Then the identity operation is the desired. Now we
may suppose $\psi_2\neq 0$. Let $A=\left(\begin{array}{cc}a&
1-a\\1-a&a\end{array}\right)$ ($0\leq a\leq 1$) be the doubly
stochastic matrix such that \begin{equation}\left(\begin{array}{c}\psi_1^2\\
\psi_2^2\end{array}\right)=\left(\begin{array}{cc}a&
1-a\\1-a&a\end{array}\right)\left(\begin{array}{c}\phi_1^2\\
\phi_2^2\end{array}\right). \end{equation}Define
\begin{equation}K_1=\left(\begin{array}{cc}\sqrt a \frac{\phi_1}{\psi_1}&
0\\0&\sqrt {a}
\frac{\phi_2}{\psi_2}\end{array}\right),\end{equation}
\begin{equation}K_2=\left(\begin{array}{cc} 0&\sqrt {1-a} \frac{\phi_1}{\psi_2}
\\ \sqrt {1-a} \frac{\phi_2}{\psi_1}&0\end{array}\right).\end{equation} One
can check that the incoherent operation whose Kraus operators are
$K_1,K_2$ is the desired.

Assume the result holds true for $\dim H\leq d-1$, we will prove
that the result holds true for $\dim H=d$ and divide the proof
into two cases.

Case 1. There is a $k$ ($  1< k< d$) such that $\psi_k\neq 0$ and
$\psi_{k+1}=\cdots=\psi_d=0$. From the majorization, it follows
that $\phi_{k+1}=\cdots=\phi_d=0$. The $k$ level vector
$(|\psi_1|^2, \cdots, |\psi_k|^2)^t$ is majorized by $(|\phi_1|^2,
\cdots, |\phi_k|^2)^t$. From the inductive assumption, there is an
incoherent operation $\widetilde{\Phi}$ on $M_k$ (the set of all
$k\times k$ level matrices) specified the Kraus operators
$\widetilde{K_n}$($n=1,2,\cdots,N$) such that $\sum_{i=1}^k
\psi_i|i\rangle\xrightarrow{\widetilde{\Phi}}\sum_{i=1}^k
\phi_i|i\rangle$. Define $K_n=\widetilde{K_n}\oplus \frac 1 {\sqrt
N}I_{d-k}$, then $\Phi(\cdot)=\sum_{i=1}^N K_n\cdot K_n^\dag$ is
an incoherent operation which transforms
$|\psi\rangle\langle\psi|$ to $|\phi\rangle\langle\phi|$.

Case 2. $\psi_d\neq 0$. Let $A$ be a doubly stochastic matrix with
$(|\psi_1|^2,\cdots, |\psi_d|^2)^t=A(|\phi_1|^2,\cdots,
|\phi_d|^2)^t$. Note that the composition of incoherent operations
are also incoherent, by the fact (ii),    $A$ can be reduced to a
$T-$transform for some indices $i,j$ and $0\leq t\leq 1$. Let
$\pi=(1,2,\cdots,i-1,j,i+1,\cdots, j-1, i,j+1,\cdots, d)$ be a
permutation of $\{1,2,\cdots,d\}$, and
\begin{equation}K_1=\sqrt{t}\text{diag}(\frac{\phi_1}{\psi_{1}},\cdots,\frac{\phi_d}{\psi_{d}}),\end{equation}
\begin{equation}\begin{array}{ll}
K_2= &\sqrt{1-t}\text{diag}(\frac{\phi_1}{\psi_{1}},\cdots,
\frac{\phi_{i-1}}{\psi_{i-1}},
\frac{\phi_i}{\psi_{j}},\frac{\phi_{i+1}}{\psi_{i+1}},\cdots,\\&
\frac{\phi_{j-1}}{\psi_{j-1}},\frac{\phi_j}{\psi_{i}},
\frac{\phi_{j+1}}{\psi_{j+1}},\cdots,
\frac{\phi_d}{\psi_{d}})P_{\pi}.\end{array}\end{equation} Then
\begin{equation}K_1^{\dag}K_1=t\text
{diag}(\frac{\phi_1^2}{\psi_{1}^2},\cdots,\frac{\phi_d^2}{\psi_{d}^2}),\end{equation}
\begin{equation}\begin{array}{ll}
K_2^{\dag}K_2=
&(1-t)\text{diag}(\frac{\phi_1^2}{\psi_{1}^2},\cdots,
\frac{\phi_{i-1}^2}{\psi_{i-1}^2},
\frac{\phi_j^2}{\psi_{i}^2},\frac{\phi_{i+1}^2}{\psi_{i+1}^2},\cdots,\\&
\frac{\phi_{j-1}^2}{\psi_{j-1}^2},\frac{\phi_i^2}{\psi_{j}^2},
\frac{\phi_{j+1}^2}{\psi_{j+1}^2},\cdots,
\frac{\phi_d^2}{\psi_{d}^2}).\end{array}\end{equation} From
$(|\psi_1|^2,\cdots, |\psi_d|^2)^t=A(|\phi_1|^2,\cdots,
|\phi_d|^2)^t$, it follows that $K_1^{\dag}K_1+K_2^{\dag}K_2=I$.
Furthermore, it is easy to check that
$\Phi(\cdot)=\sum_{n=1}^2K_n\cdot K_n^{\dag}$ transforms
$|\psi\rangle\langle\psi|$ to $|\phi\rangle\langle\phi|$. Note
that each column of $K_n(n=1,2)$ has at most one nonzero entry, so
$\Phi$ is incoherent. This finishes the proof of the ``if" part.

To prove the converse, we only consider the three dimensional case,
other cases can be treated similarly. Now, we suppose $\dim H=3$
and there is an incoherent operation $\Phi$ transforms
$|\psi\rangle\langle\psi|$ to $|\phi\rangle\langle\phi|$.  Let
\begin{equation}\Phi(|\psi\rangle\langle \psi|)=\sum_n K_n|\psi\rangle\langle
\psi|K_n^\dag=|\phi\rangle\langle \phi|.\end{equation}  Hence
there exist complex numbers $\alpha_n$ such that
$K_n|\psi\rangle=\alpha_n|\phi\rangle$. Let $k_j^{(n)}$($j=1,2,3$)
be the nonzero element of $K_n$ at $j-th$ column (if there is no
nonzero element in $j-th$ column, then $k_j^{(n)}=0$). Suppose
$k_j^{(n)}$ locates $f_n(j)-th$ row. Here, $f_n(j)$ is a function
that maps  $\{2,3\}$ to  $\{1,2,3\}$ with the property that $1\leq
f_n(j)\leq j$. Let
            $\delta_{s,t}=\left\{\begin{array}{cc}
            1,&s=t\\
            0,&s\neq  t\end{array}\right.$. Then there
is a permutation $\pi_n$ such that
\begin{equation}K_n=P_{\pi_n}\left(\begin{array}{ccc}
            k_1^{(n)} & \delta_{1,f_n
            (2)}k_2^{(n)} & \delta_{1,f_n(3)}k_3^{(n)}\\
            0         & \delta_{2,f_n(2)}k_2^{(n)} & \delta_{2,f_n(3)}k_3^{(n)}\\
            0         &    0                     &
            \delta_{3,f_n(3)}k_3^{(n)}\end{array}\right).\end{equation}
From $\sum_n K_n^{\dag}K_n=I$, we get that
\begin{equation}\left\{\begin{array}{l}
\sum_n |k_j^{(n)}|^2=1, (j=1,2,3),\\
\sum_n \overline{k_1^{(n)}}\delta_{1,f_n(2)}k_2^{(n)}=0,\\
\sum_n \overline{k_1^{(n)}}\delta_{1,f_n(3)}k_3^{(n)}=0,\\
\sum_n
(\delta_{1,f_n(2)}\delta_{1,f_n(3)}+\delta_{2,f_n(2)}\delta_{2,f_n(3)})\overline{k_2^{(n)}}k_3^{(n)}=0.
\end{array}\right.\end{equation}
For $|\psi\rangle =(\psi_1,\psi_2,\psi_3)^t$, by a direct
computation, one can get
\begin{equation}K_n|\psi\rangle=P_{\pi_n}\left(\begin{array}{c}
k_1^{(n)}\psi_1 + \delta_{1,f_n(2)}k_2^{(n)}\psi_2 +
\delta_{1,f_n(3)}k_3^{(n)}\psi_3\\
\delta_{2,f_n(2)}k_2^{(n)}\psi_2+ \delta_{2,f_n(3)}k_3^{(n)}\psi_3\\
\delta_{3,f_n(3)}k_3^{(n)}\psi_3\end{array}\right),\end{equation}
and so
\begin{equation}\left\{\begin{array}{l}
k_1^{(n)}\psi_1 + \delta_{1,f_n(2)}k_2^{(n)}\psi_2 +
\delta_{1,f_n(3)}k_3^{(n)}\psi_3=\alpha_n \phi_{\pi_n^{-1}(1)},\\
\delta_{2,f_n(2)}k_2^{(n)}\psi_2+ \delta_{2,f_n(3)}k_3^{(n)}\psi_3=\alpha_n \phi_{\pi_n^{-1}(2)},\\
\delta_{3,f_n(3)}k_3^{(n)}\psi_3=\alpha_n
\phi_{\pi_n^{-1}(3)}.\end{array}\right.\end{equation} Applying
$\sum_n|\cdot|^2$ to above equations, we have
\begin{equation}\left\{ \begin{array}{l}
\psi_1^2 + \sum_n\delta_{1,f_n(2)}|k_2^{(n)}|^2\psi_2^2 \\
\ \ \ \ +
\sum_n\delta_{1,f_n(3)}|k_3^{(n)}|^2\psi_3^2\\
\ \ \ \ +\sum_n\delta_{1,f_n(2)}\delta_{1,f_n(3)}\overline{k_2^{(n)}}k_3^{(n)}\psi_2\psi_3\\
\ \ \ \
+\sum_n\delta_{1,f_n(2)}\delta_{1,f_n(3)}\overline{k_3^{(n)}}k_2^{(n)}\psi_3\psi_2\\
\ \ =
\sum_n|\alpha_n |^2\phi_{\pi_n^{-1}(1)}^2,\\
\sum_n\delta_{2,f_n(2)}|k_2^{(n)}|^2|\psi_2^2+
\sum_n\delta_{2,f_n(3)}|k_3^{(n)}|^2|\psi_3^2\\
\ \ \  \ +\sum_n\delta_{2,f_n(2)}\delta_{2,f_n(3)}\overline{k_2^{(n)}}k_3^{(n)}\psi_2\psi_3\\
\ \ \ \ +\sum_n\delta_{2,f_n(2)}\delta_{2,f_n(3)}\overline{k_3^{(n)}}k_2^{(n)}\psi_3\psi_2\\
\ \ =\sum_n|\alpha_n |^2\phi_{\pi_n^{-1}(2)}^2 ,\\
\sum_n\delta_{3,f_n(3)}|k_3^{(n)}|^2|\psi_3^2=\sum_n|\alpha_n
|^2\phi_{\pi_n^{-1}(3)}^2.\end{array}\right.\end{equation} Note
that, for $s=1,2,3$,
\begin{equation}\begin{array}{ll}
&\sum_n|\alpha_n |^2\phi_{\pi_n^{-1}(s)}^2\\
=& \sum_{n,\pi_n^{-1}(s)=1}|\alpha_n
|^2\phi_1^2+\sum_{n,\pi_n^{-1}(s)=2}|\alpha_n
|^2\phi_2^2\\
& +\sum_{n,\pi_n^{-1}(s)=3}|\alpha_n
|^2\phi_3^3,\end{array}\end{equation} Let
$d_{ij}=\sum_{n,\pi_n^{-1}(i)=j}|\alpha_n |^2, 1\leq i,j\leq 3$,
then the matrix $D=(d_{ij})$
 is a doubly stochastic
matrix, since $\sum_{n}|\alpha_n |^2=1$. Furthermore,
\begin{equation}\begin{array}{ll}&D(\phi_1^2,\phi_2^2,\phi_3^2)^t\\
=&(\sum_n|\alpha_n |^2\phi_{\pi_n^{-1}(1)}^2,\sum_n|\alpha_n
|^2\phi_{\pi_n^{-1}(2)}^2,\sum_n|\alpha_n
|^2\phi_{\pi_n^{-1}(3)}^2)^t.\end{array}\end{equation} This
implies that
\begin{equation}\begin{array}{ll}&(\sum_n|\alpha_n |^2\phi_{\pi_n^{-1}(1)}^2,\sum_n|\alpha_n
|^2\phi_{\pi_n^{-1}(2)}^2,\sum_n|\alpha_n
|^2\phi_{\pi_n^{-1}(3)}^2)^t\\ \prec &
(\phi_1^2,\phi_2^2,\phi_3^2)^t.\end{array}\end{equation} On the
other hand, note that in equation (21), \begin{equation}\begin{array}{l}\sum_n\delta_{1,f_n(2)}|k_2^{(n)}|^2\psi_2^2 \\
\ \ \ \ +
\sum_n\delta_{1,f_n(3)}|k_3^{(n)}|^2\psi_3^2\\
\ \ \ \ +\sum_n\delta_{1,f_n(2)}\delta_{1,f_n(3)}\overline{k_2^{(n)}}k_3^{(n)}\psi_2\psi_3\\
\ \ \ \
+\sum_n\delta_{1,f_n(2)}\delta_{1,f_n(3)}\overline{k_3^{(n)}}k_2^{(n)}\psi_3\psi_2\\
=\sum_n
|\delta_{1,f_n(2)}k_2^{(n)}\psi_2+\delta_{1,f_n(3)}k_3^{(n)}\psi_3|^2.\end{array}\end{equation}
From the definition of majorization and equations (18),(21), one can
check that
\begin{equation}\begin{array}{ll}
&(\psi_1^2,\psi_2^2,\psi_3^2)^t\\
\prec & (\sum_n|\alpha_n |^2\phi_{\pi_n^{-1}(1)}^2,\sum_n|\alpha_n
|^2\phi_{\pi_n^{-1}(2)}^2,\sum_n|\alpha_n
|^2\phi_{\pi_n^{-1}(3)}^2)^t.\end{array}\end{equation} Therefore
$(\psi_1^2,\psi_2^2,\psi_3^2)^t\prec
(\phi_1^2,\phi_2^2,\phi_3^2)^t$.

{\it Outlook.---}Our results raise many interesting questions. It
would be of great interest to determine when a mixed state $\rho$
can be transformed to a mixed state $\sigma$ by incoherent
operations. What we get is if $\sigma$ is incoherent then there
exists an incoherent operation $\Phi$ such that
$\Phi(\rho)=\sigma$ for any state $\rho$. We show this by
explicitly constructing an incoherent operation that achieves the
transformation in the Appendix. What are sufficient conditions for
the existence of catalysts? Finally, all of considerations above
implicitly assumed the finite dimensional setting, but this is
neither necessary nor desirable as there are very relevant
physical situations that require infinite dimensional systems for
their description. Most notable are the quantum states of light,
that is quantum optics, with its bosonic character requires
infinite dimensional systems, harmonic oscillators, for their
description. Hence, coherence manipulation and existence of
catalysts  in infinite dimensional systems are needed. Mirroring
analogous developments in entanglement manipulation \cite{Mas}, we
expect that the manipulation of coherence in  infinite dimensional
systems  can be built.

{\it Conclusions.---} In this manuscript, we give a complete
 characterization of  coherence manipulation for pure states in
 terms of majorization. This result offers an affirmative
answer to the open question: whether, given two states $\rho$ and
$\sigma$, either $\rho$ can be transformed into $\sigma$ or vice
versa under incoherent operations \cite{Bau}. The proof of the
result also provides a effective constructive method to find the
incoherent operation transforming $|\psi\rangle$  to
$|\phi\rangle$, whenever $(|\psi_1|^2,\cdots,
|\psi_d|^2)^t\prec(|\phi_1|^2,\cdots, |\phi_d|^2)^t$. The
majorization approach used here is similar to that used to
establish the ordering of entanglement states, which led to
advancement in the field of quantum computation. Based on Theorem
1, some interesting properties of coherent catalysts are
discovered.

{\it Acknowledgement.---} The authors thank referees for their
valuable comments which improve the presentation of this paper.
This work was completed while the authors were visiting the
Institute of Quantum Computation of the University of Waterloo
during the academic year 2014-2015 under the support of China
Scholarship Council. We thank Professor David W. Kribs and
Professor Bei Zeng for their hospitality. This work is partially
supported by the Natural Science Foundation of China (No.
11001230, 11301312, 11171249), the Natural Science Foundation of
Fujian (2013J01022, 2014J01024), the Natural Science Foundation of
Shanxi ( No. 2013021001-1, 2012011001-2) and the Research start-up
fund for Doctors of Shanxi Datong University under Grant No.
2011-B-01.

\begin{center} \textbf {Appendix : Transition of mixed states}\end{center}

We will show if the output mixed state $\sigma$ is incoherent,
i.e., $\sigma\in {\mathcal I}$, then for any quantum state $\rho$,
there exists an incoherent operation $\Phi$ such that
$\Phi(\rho)=\sigma$. We do this by an explicit construction of an
incoherent operation. Define an incoherent operation
$$\Phi_1(\rho):=\sum_{i=1}^{d}|i\rangle\langle i|\rho|i\rangle \langle i|.$$
The effect of this operation is to remove all off-diagonal
elements of $\lambda_{i,j}|i\rangle\langle j|(i\neq j)$ from
$\rho=\sum_{i,j=1}^{d}\lambda_{i,j}|i\rangle\langle j|$, leaving
the diagonal elements $\lambda_{i,i}|i\rangle\langle i|$ intact.
Denote $\{\lambda_i=\langle i|\rho |i\rangle\}_{i=1}^{d}$ and
$\{\mu_i\}_{i=1}^{d}$  the eigenvalues of $\Phi_1(\rho)$ and
$\sigma$, respectively. Let
$$\begin{array}{ll}
A_1=&\sqrt{\mu_1}|1\rangle\langle 1|+\sqrt{\mu_2}|2\rangle\langle
2|+\cdots+\sqrt{\mu_d}|d\rangle\langle d|,\\
A_2=&\sqrt{\mu_2}|1\rangle\langle 2|+\sqrt{\mu_3}|2\rangle\langle
3|+\cdots\\
& +\sqrt{\mu_{d}}|d-1\rangle\langle d|+\sqrt{\mu_{1}}|d\rangle\langle 1|,\\
\cdots,&\\
A_i=&\sqrt{\mu_i}|1\rangle\langle
i|+\cdots+\sqrt{\mu_{m_{s+i-1}}}|s\rangle\langle
m_{s+i-1}|+\cdots\\
& +\sqrt{\mu_{m_{d+i-1}}}|d\rangle\langle
\mu_{m_{d+i-1}}|,\\
\cdots,& \\
A_d=&\sqrt{\mu_d}|1\rangle\langle d|+\sqrt{\mu_1}|2\rangle\langle
1|+\cdots+\sqrt{\mu_{d-1}}|d\rangle\langle d-1|,
\end{array}$$ here $m_x=x-[\frac{x-1}{d}]d$. It is easy to check
that $\sum_{i=1}^dA_iA_i^\dag=I$. By a direct computation, one can
get $\Phi_2(\Phi_1(\rho))=\sum_{i=1}^d
A_i^\dag\Phi_1(\rho)A_i=\sigma$. Let $\Phi=\Phi_2\circ\Phi_1$,
then $\Phi$ is an incoherent operation satisfying
$\Phi(\rho)=\sigma$.


\begin{thebibliography} {99}

\bibitem{Nielsen} M. A. Nielsen and I. L. Chuang, Quantum Computation and Quantum
information (Cambridge University Press, Cambridge, 2000).

\bibitem{Ben1} C.H. Bennett et al., Phys. Rev. A \textbf{54}, 3824
(1996).

\bibitem{Ben2} C.H. Bennett et al., Phys. Rev. A \textbf{53}, 2046
(1996).

\bibitem{Ben3} C.H. Bennett et al., Phys. Rev. Lett. \textbf{76},
722 (1996).

\bibitem{Ved1} V. Vedral et al., Phys. Rev. Lett. \textbf{78},
2275 (1997).

\bibitem{Ved2} V. Vedral and M.B. Plenio., Phys. Rev. A \textbf{57},
1619 (1998).



\bibitem{Nie2} M.A. Nielsen, Phys. Rev. Lett. \textbf{83}, 436 (1999).

\bibitem{Vid} G. Vidal, Phys. Rev. Lett. \textbf{83}, 1046 (1999).

\bibitem{Jon1} D. Jonathan and M.B. Plenio,
Phys. Rev. Lett. \textbf{83}, 1455 (1999).

\bibitem{Jon2} D. Jonathan and M.B. Plenio,
Phys. Rev. Lett. \textbf{83}, 3566 (1999).

\bibitem{Har} L. Hardy,
Phys. Rev. A \textbf{60}, 1912 (1999).

\bibitem{Nie3}  M.A. Nielsen, G. Vidal, Quantum Inf. Comput.
\textbf{1}, 76 (2001).

\bibitem{Bau} T. Baumgratz, M. Cramer, M.B. Plenio,
Phys. Rev. Lett.  \textbf{113},  140401 (2014).


\bibitem{Bha} R. Bhatia, Matrix Analysis (Springer-Verlag, New York, 1997).


\bibitem{Ghi} I. Ghiu, M. Bourennane, A. Karlsson, Phys. Lett. A
\textbf{287}, 12 (2001).

\bibitem{Son} W. Song, Y. Huang, N.L. Liu, Z.B. Chen, J. Phys. A:
Math. Gen. \textbf{40}, 785 (2007).

\bibitem{Cha} I. Chattopadhyay, D. Sarkar, Phys. Rev. A
\textbf{77}, (2008) 05305(R).

\bibitem{Bow} G. Bowen, N. Datta, IEEE Trans. Inf. Theory
\textbf{54}, 3677 (2008).

\bibitem{Ber} J. \AA berg, Phys. Rev. Lett. \textbf{113}, 150402 (2014).


\bibitem {Gir} D. Girolami, Phys. Rev. Lett. \textbf{113}, 170401 (2014).

\bibitem{BDQ} S.-P. Du, Z.-F. Bai, X.-F. Qi, 
arXiv:1504.02862v1.

\bibitem{Mas} M. Owari, S.L. Braunstein, K. Nemoto, M. Murao,
Quantum Inf. Comput. \textbf{8}, 30 (2008).




\end{thebibliography}
\end{document}